\documentclass[11pt]{article}
\def\beqar {\begin{eqnarray}}
\def\eeqar {\end{eqnarray}}
\def\beq {\begin{equation}}
\def\eeq {\end{equation}}
\def \half {{\textstyle{1\over 2}}}
\def\vf {{\varphi}}
\def\Tr {{\rm Tr}}

\def\del {\partial}

\def\half {{\textstyle {1 \over 2}}}
\def\vf {\varphi}
\def\ra {\rangle}
\def\la {\langle}
\def\Tr {{\rm Tr}}

\def \C {{\cal C}}
\def \D {{\cal D}}

\def \G {{\cal G}}
\def \A {{\cal A}}

\begin{document}

\begin{titlepage}
\null\vspace{-62pt}

\pagestyle{empty}
\begin{center}
\rightline{CCNY-HEP-03/1}

\vspace{1.0truein} {\Large\bf On the invariant measure for the
Yang-Mills}\\
\vskip .1in
{\Large \bf configuration space in (3+1) dimensions} 

\vspace{.5in}V.P. NAIR\\
\vspace{.1in}{\it Physics Department\\
City College of the CUNY\\  
New York, NY 10031}\\
E-mail: vpn@sci.ccny.cuny.edu\\
\vspace {.3in} 
A. YELNIKOV\\
\vskip .1in{\it Physics Department\\
Queens College of the CUNY\\ 
Flushing, NY 11367}\\
E-mail: yelnikov@qcunix1.qc.edu\\
\vspace{0.1in}
\end{center}
\vspace{0.5in}

\centerline{\bf Abstract}
We consider a gauge-invariant Hamiltonian analysis for Yang-Mills 
theories in three spatial
dimensions. The gauge potentials are parametrized in terms of
a matrix variable which facilitates the elimination of the
gauge degrees of freedom. We develop an approximate calculation of
the volume element on the gauge-invariant configuration space. 
We also make a rough estimate of the ratio of $0^{++}$ glueball mass
and the square root of
string tension by comparison with $(2+1)$-dimensional Yang-Mills
theory.
\baselineskip=18pt

\end{titlepage}

\hoffset=0in
\newpage
\pagestyle{plain}
\setcounter{page}{2}
\newpage

\section{Introduction}

In a series of recent papers a Hamiltonian analysis of Yang-Mills
theories in (2+1) dimensions was developed \cite{KN, KKN1, KKN2}.
This was mainly motivated by the fact that, 
while it is true that gauge theories of direct
physical interest are in (3+1) dimensions,
the study of Yang-Mills gauge
theories in two spatial dimensions can be useful for two reasons. It can
be a guide to the more realistic case of three dimensions, and secondly,
gauge theories in two spatial dimensions can be interpreted as 
an approximation
to the  high-temperature phase of QCD with the mass gap playing the role of 
the magnetic
mass. (It should be pointed out that, precisely for these reasons, 
there have been
many analyses of (2+1)-dimensional gauge theories starting 
from the early days \cite{everybody}.) In this paper, 
we shall start a similar Hamiltonian analysis
of Yang-Mills theories in (3+1) dimensions, 
carrying over some of the lessons from
the lower dimensional analysis.

In the (2+1) dimensional theory, the $A_0=0$ gauge was chosen and
the complex
components of the spatial gauge field, viz.,
$A_z,~A_{\bar z}$ were parametrized as
$A_z = -\partial_z M~M^{-1}$, $A_{\bar z} =M^{\dagger -1}\partial_{\bar
z} M^\dagger$, where $M, M^\dagger$ are $SL(N, {\bf C})$-matrices
for an $SU(N)$ gauge theory. 
The basic gauge-invariant variable for the theory is then
the hermitian
matrix field 
$H=M^\dagger M$. This particular parametrization of the potentials proved to be
very useful since the  Jacobian for the transformation of variables and
the volume element on space of gauge-invariant configurations could be
exactly calculated. 
This invariant volume measure on the physical configuration space,
which also determines the inner products
for wavefunctions,
is given  in terms of the
Wess-Zumino-Witten (WZW) action for the field $H$ \cite{witten, gaw, KN}. 
Considerations of integrable representations
of the WZW model then showed that normalizable wavefunctions are
functions of the current
$J= (N/\pi ) \partial_zH~H^{-1}$. In other words, 
the wavefunctions have to be more
restricted than being just functions of
$H$; they can only depend on $H$ via the specific combination in $J$. 
The regularized
kinetic energy operator, which is the Laplacian on this 
infinite-dimensional
configuration space, is given
in terms of functional derivatives with respect to $J$; 
the potential energy can also be written in terms of $J$ \cite{KKN1}.
The vacuum wavefunction $\Psi_0$ of the theory was obtained by solving the
(functional) Schr\"odinger equation in the approximation of  keeping all
terms in
$\log \Psi_0$ which are quadratic in $J$, with a  
systematic expansion for the
higher order terms. 
The vacuum wavefunction agrees with perturbation
theory for the high momentum modes.
The expectation value of
the Wilson loop operator and hence the string tension were calculated
\cite{KKN2}. 
The values
for the string tension agree within $3\%$ of recent Monte Carlo 
evaluations \cite{teper}.
Finally, the propagating
particles in the perturbative regime can be shown to
have a mass $m=e^2N/2\pi$.  
This may be taken as a prediction for the magnetic mass
of gluons in high temperature QCD  \cite{VPN1}.
This result compares favorably with resummation calculations of this
quantity \cite{magmass} and with lattice
estimates, keeping in mind that this is a difficult lattice calculation as
well \cite{karsch}. 
Finally, these techniques can also be
extended to the Yang-Mills-Chern-Simons
theory \cite{KKN3}. 

While this Hamiltonian analysis still leaves
many open questions, it is fair to claim that
some progress  in understanding the (2+1)-dimensional case has indeed been achieved.
It is worth noting that the vacuum
wavefunction which was obtained, irrespective of the calculations preceding it, has the
desirable features of agreeing with the perturbative vacuum wavefunction in 
the high momentum
limit and giving an area law for the Wilson loop with a
string tension which agrees closely with the lattice calculations.
Therefore, further study along these lines, in particular
exploring a similar strategy in (3+1) dimensions, is warranted.

In section 2, we will introduce the parametrization of the gauge potentials
in terms of the matrix variables. The calculation of the volume measure 
of the configuration space (and hence the inner product for
wavefunctions) is taken up in section 3.
Section 4 gives some remarks on this result and,
by comparison with $(2+1)$-dimensional Yang-Mills theory,
makes a rough estimate of the ratio of $M_{0^{++}}/\sqrt{\sigma}$
where $M_{0^{++}}$ is the mass of the $0^{++}$ glueball
and $\sigma$ is the
string tension.

\section{The parametrization of the gauge potentials}

We shall discuss an $SU(N)$-gauge
theory and also choose the gauge $A_0=0$, as is convenient for a 
Hamiltonian formulation.
The remaining gauge potentials can be
written as
$A_i = -i t^a A_i ^a$, $i=1,2,3$, where $t^a$ are hermitian 
$(N \times N)$-matrices which form a basis of the Lie algebra of $SU(N)$ with
$[t^a, t^b ] = i f^{abc} t^c,~~{\rm {Tr}} (t^at^b) = {1 \over 2} \delta ^{ab}$. 

We start by recalling that the key ingredients of the
(2+1) dimensional analysis were the following:
\begin{enumerate}
\item the parametrization of the potentials in terms of the matrix $M$ which allowed
the realization of gauge transformations in a homogeneous way, $M^g =gM$.
\item the calculation of the gauge-invariant measure on the configuration space.
\item evaluation of the Hamiltonian in terms of these gauge-invariant variables.
\item solving the functional Schr\"odinger equation for the vacuum wave function.
\item calculating the string tension and other quantities of interest.
\end{enumerate}

The study of the first two steps in (3+1) dimensions will be taken up in this paper.
Let ${\A}$ denote the set of all gauge potentials $A_i ^a$. 
Gauge transformations act
on $A_i$ in the standard way, $A_i \rightarrow A_i ^g$, where
\beq
A_i^{(g)} = g A_i g^{-1} - \partial_i g g^{-1} 
\label{1}
\eeq
and $g( \vec x ) \in SU(N)$. The gauge group ${\G}_*$ is defined by
\beq
{\G}_* = \left\{ {\rm set ~of ~all} ~g(\vec x ): {\bf R}^3 \rightarrow SU(N),
~~g\rightarrow 1 ~{\rm as} ~\vert \vec x \vert \rightarrow \infty \right\}
\label{2}
\eeq
The space of gauge-invariant field configurations is $\C ={\A}/{\G}_*$.
A parametrization of the gauge potentials is equivalent to choosing 
coordinates on the configurations space. Since the space $\C$ has 
nontrivial topology, any parametrization is restricted to some
open region. We use a parametrization in a region which includes
$A=0$ and calculate (approximately) the volume measure of $\C$
for this region. (Not surprisingly, the geometry and topology of 
the Yang-Mills
configuration space in three spatial dimensions have also been
studied by a number of authors, see references \cite{singer, mitter}.
For a recent summary and new results on the metric, see
\cite{orland}.)

Going back to $YM_{2+1}$, we start by asking why it is possible to parametrize
$A_z $ as $-\del_z M M^{-1}$. Notice that this parametrization
may be written as $(\del_z +
A_z )M =0$ and one can convert it to an integral equation
\beqar
M (x) &=& 1 - \int_{x'}  S(x, x')~ A_z (x') M(x')\nonumber\\
\del_z ~S(x, x')&=& \delta^{(2)}(x -x')
\label{3}
\eeqar
With this equation, we see that, at least iteratively, 
we can find an $M$ for each given
$A_z$. This establishes a mapping $A_z \rightarrow M$. 
(There are much more elegant
and more general ways to justify the parametrization $A_z=- \del_z M M^{-1}$,
but this simple argument is most suitable for what follows 
\cite{KN}.) Notice that the key 
ingredient is the invertibility of $\del_z$. The first
term involving $A$ in a series expansion for $M$, namely,
$\int (\del_z)^{-1} A_z$, is a complex matrix which is traceless since 
$A_z$ has no trace. It is thus an element of the Lie algebra of $SL(N,{\bf C})$,
showing that $M$ can be taken to be in $SL(N,{\bf C})$. Conversely,
$M$ contains ${\rm dim} [SL(N,{\bf C})] = 2\times {\rm dim} [SU(N)]$ independent functions
corresponding exactly to the number of independent functions needed
for the potential, $A_i,~i=1,2$, therefore one has the map $M\rightarrow A_z$ as well.

Since $\del_z$ is the chiral Dirac operator in two dimensions,
the invertibility of $\del_z$ is equivalent to the existence of the propagator
for the chiral Dirac theory. In three Euclidean dimensions, which
is appropriate for the (3+1)-dimensional theory, there is no chirality, but
we can use the Dirac operator $\sigma \cdot \del$ where
$\sigma_i,~ i=1,2,3,$ are the Pauli matrices. We then {\it define}
a matrix $M$ by 
\beq
\left( \sigma\cdot \del + \sigma\cdot A \right) M =0
\label{4}
\eeq
On such a matrix $M$, gauge transformations act by
$M \rightarrow M^g = g M$, where $g$ is an element of
$SU(N)$.
Equation (\ref{4}) has the formal inversion
\beq
M (x) = 1 - \int_y \left( {1\over \sigma \cdot \del} \right)_{xy} 
\sigma\cdot A(y)~ M(y) 
\label{5}
\eeq
where
\beqar
\left( {1\over \sigma \cdot \del} \right)_{xy}&=&
- \int {d^3p \over (2\pi )^3} ~{i\sigma\cdot p \over p^2}~ 
e^{ip\cdot (x-y)}\nonumber\\
&=& -\sigma\cdot \del_x G(x,y)\nonumber\\
G(x,y)&=&\int {d^3p \over (2\pi )^3} ~{1\over p^2}~ 
e^{ip\cdot (x-y)}
\label{6}
\eeqar
To first order in the $A$'s, the solution for $M$ is then
\beqar
M &\approx& 1 -it^a \theta^a +
\sigma_i t^a \int_y G(x,y) \epsilon_{ijk} \del_j A^a_k(y)\nonumber\\
\theta^a &=& \int_y G(x,y) \del\cdot A^a (y) 
\label{7}
\eeqar
The term $-it^a \theta^a$
on the right hand side of expression (\ref{7}) for $M$
can be removed by a
gauge transformation of the form
$M \rightarrow \exp ( it^a \theta^a)~M$, consistent with the fact that
$\del\cdot A$ represents the gauge degree of freedom, to linear order
in $A$.
The last term shows that the infinitesimal generators,
for whatever group $M$ belongs to, must include
$\sigma_i t^a$, which are a subset of generators
of  $SU(2N, {\bf C})$.
Completion of the algebra under commutation
rules shows that we need all of $SU(2N, {\bf C})$. 
Thus generally we must take $M$ to be an element
of $SU(2N, {\bf C})$. Equation (\ref{4}) thus gives a map $A\rightarrow M
\in SU(2N, {\bf C})$.

An arbitrary element of $SU(2N, {\bf C})$ will contain
$2\times (4N^2 -1)$ parameter functions. Thus arbitrary 
$SU(2N, {\bf C})$-matrix functions
$M$ contain too many parameters to give a faithful coordinatization
of a region of $\A$,
we will need to
use constraints on $M$. We will now work out the required constraints.
For most of what follows, it is convenient to use 
$U(2N)$ rather than $SU(2N)$. We define the set of hermitian matrices
$\{ t^A \},~A=1,2,\cdots,4N^2$ as the set $\{ 1\otimes t^a, \sigma_i \otimes t^a\}$,
$a= 1,2,\cdots, N^2$. $t^a$ are taken as $(N\times N)$ hermitian
matrices
normalized by $\Tr t^a t^b = \half \delta^{ab}$. $\{ -it^a\}$ form an antihermitian
basis for $U(N)$ embedded in $U(2N)$. The set of matrices
$\{ -it^A, t^A\}$ form a basis for
the Lie lagebra of $U(2N, {\bf C})$.
The normalization condition for the $t^A$ is $\Tr t^A t^B = \delta^{AB}$;
they further obey the completeness relation $t^A_{mn} t^A_{pq}
= \delta_{np} \delta_{mq}$. 

Now let $M$ be an arbitrary $U(2N, {\bf C})$ matrix. We can then expand
\beq
\sigma \cdot \del M M^{-1} = i ~\phi^a t^a + i~\sigma \cdot A^a t^a
\label{8}
\eeq
where $\phi^a, A^a_i$ are in general complex functions. If we were to start 
from real $A^a_i$ and
use equation (\ref{4}), then $\phi^a$ in (\ref{8}) would be zero. 
Thus, to eliminate unwanted degrees of freedom starting from
an arbitrary $M$, we must impose the conditions
$A^a_i - {\bar A}^a_i =0,~~\phi^a =0$. These are equivalent to
\beqar
\Tr (t^a \sigma \cdot \del M M^{-1} ) &=& 0 \label{9}\\
\del_i \cdot (M^\dagger \sigma_i M) &=& 0
\label{10}
\eeqar
The only remaining degree of freedom in $M$ then corresponds to
the real part of $A^a_i$ which is the $U(N)$-gauge potential.

It is instructive to work out these conditions for $M$ close to the identity.
Writing $M\approx 1 + it^a \vf^a +i \sigma_i t^a \Theta^a_i$, we find
\beq
\sigma\cdot \del M M^{-1} = it^a \del_i \Theta^a_i ~+~ i \sigma_k t^a \left( \del_k \vf^a +i
\epsilon_{ijk}
\del_i \Theta^a_j\right)
\label{11}
\eeq
Imposing the constraint (\ref{10}) on this, and separating out the real 
and imaginary parts
of the functions, we get
\beqar
\del_i ({\rm Im} \Theta^a_i )&=& 0\nonumber\\
\del_i ({\rm Im} \vf^a ) + \epsilon_{ijk} \del_j {\rm Re} \Theta^a_k &=& 0
\label{12}
\eeqar
The second of these equations gives the Laplace equation
for ${\rm Im}\vf^a$, namely, $\del^2 {\rm Im} \vf^a =0$, so that
with proper boundary conditions, we can take ${\rm Im} \vf^a =0$.
Further, we find ${\rm Re} \Theta^a_i = \del_i \xi^a$ for some 
scalar functions $\xi^a$. Putting this back into (\ref{11}) and comparing
with (\ref{8}) we find
\beqar
A^a_i &=& \del_i \vf^a ~-\epsilon_{ijk} \del_j {\rm Im}\Theta^a_k
\nonumber\\
\phi^a &=& \del^2 \xi^a \label{13}
\eeqar 
The constraint (\ref{9}) eliminates $\phi^a$ (or $\xi^a$).
The functions $\vf^a$ (which are
now real) represent the gauge degrees of freedom. The gauge 
invariant degrees of freedom
are given by ${\rm Im} \Theta^a_i$, which are only
two polarizations ($2\times N^2$ functions) because of the 
condition $\del_i ({\rm Im} \Theta^a_i )= 0$ in (\ref{12}).
It is rather well known that an Abelian gauge potential
can be parametrized in the form given in (\ref{13}),
$A_i = \del_i \vf ~-\epsilon_{ijk} \del_j \Theta_k$
with $\del_i \Theta_i =0$.
Near the zero potential, a similar parametrization will
apply to the $U(N)$ potentials as well; equation (\ref{13}) is just this, 
with the required $N^2$ replication of the functions.

Since $M=1$ corresponds to the zero potential, the above analysis
shows that an arbitrary $U(2N, {\bf C})$-matrix $M$, subject to the
conditions (\ref{9}, \ref{10}), does give a faithful coordinatization
of $\A$ for a region containing the zero potential. For, given
any $A$, we can generate a corresponding $M$ by solving (\ref{4})
and conversely, given any $M$ subject to (\ref{9}, \ref{10}), we get
a general gauge potential with the correct number of degrees of freedom.
How much of $\A$ or $\C$ can be covered by
this parametrization is a very valid and interesting question;
at this stage there is no clear answer to this.
This is also evidently related to the question of Gribov ambiguities
and other topological issues for ${\C}$ \cite{gribov, singer,
baal}.
In the (2+1)-dimensional case,
a similar question arises for the parametrization $A_z =-\del_z M M^{-1}$.
In that case, there is an ambiguity in $M$ for a given $A$, namely,
$M$ and $MV({\bar z})$ give the same $A$; by ensuring invariance under
this holomorphic symmetry for all physical quantities, 
at least some of the difficulties of transitions from one
coordinate patch to another could be circumvented \cite{KN}.

\section{The volume measure}

We now turn to the calculation of the volume measure on the 
configuration space.
In terms of the fields $\phi^a$, $A^a_i$ given in (\ref{8}),
introduce the Euclidean metric
\beq
ds^2 =\int d^3x ~\left( \delta {\bar A}^a_i \delta A^a_i + \delta{\bar\phi}^a
\delta \phi^a \right)
\label{14}
\eeq
For the gauge potential of interest which is the real part of $A^a_i$,
this is the Euclidean metric which is precisely the metric of interest 
for the gauge theory. The Euclidean volume measure for the real part of
$A^a_i$ can be written as
\beq
[d ~{\rm Re} A^a_i] = \int [dA] ~\delta ( A^a_i - {\bar A}^a_i ) 
~\delta ( \phi^a) ~\delta ( {\bar \phi}^a)\label{15}
\eeq
$[dA]$ involves all components, $A^a_i$, ${\bar A}^a_i$ $\phi^a$ and
${\bar \phi}^a$. The functional Dirac delta functions eliminate all
except the real part of $A^a_i$. The volume $[dA]$ corresponds to the metric
(\ref{14}). From the definition (\ref{8}), we have
\beqar
\delta A^a_i &=& -i \Tr [ \sigma_i t^a \sigma_j \D_j(\delta M M^{-1})]\nonumber\\
\delta \phi^a &=& -i \Tr [ t^a \sigma_j \D_j(\delta M M^{-1})]\label{16}
\eeqar
where $\D_j$ is defined by 
\beq
\D_j \chi =
\del_j \chi + [\A_j ,\chi ], \hskip 1in \A_j = -\del_j M M^{-1}
\label{17}
\eeq
The equations in (\ref{16}) may be combined as $\delta A^A
= -i\Tr [t^A \sigma_j\D_j \theta]$,
$\theta = \delta M M^{-1}$.
Using the completeness of the $t^A$, the metric (\ref{14}) can then be
simplified as
\beqar
ds^2 &=& \int d^3x~\Tr \left({\overline {\D_i \theta }}~ \sigma_i \sigma_j~ \D_j \theta
\right)\nonumber\\
&=&\int d^3x~\Tr \left( t^A \sigma_i \sigma_j t^B\right) ({\overline {\D_i \theta }})^A
(\D_j \theta )^B
\label{18}
\eeqar
where we use the fact that, in terms of components in the Lie algebra,
$\D_j \theta = -it^A (\D_j \theta )^A$, $(\D_j \theta )^A
= \del_i \theta^A +f^{ABC} \A^B_i \theta^C$. $(T^A)_{BC} =-if^{ABC}$
are the Lie algebra generators in the adjoint representation of
$U(2N, {\bf C})$. 
We now define the $(4N^2 \times 4N^2)$- matrices
\beq
(\Sigma_i)^{AB}= \Tr (t^A \sigma_i t^B)
\label{19}
\eeq
By the completeness relation for the $t^A$, these are seen to obey
the relation
\beq
\Sigma_i^{AB} \Sigma_j^{BC} = \Tr (t^A \sigma_i \sigma_j t^C)
=\delta_{ij} \delta^{AC} +i
\epsilon_{ijk} 
\Sigma_k^{AC} 
\label{20}
\eeq
The $\Sigma_i$ are a $(4N^2 \times 4N^2)$ representation of the algebra of
$\sigma_i$. The metric (\ref{18}) can thus be written as
\beqar
ds^2 &=& \int d^3x~ ({\overline {\D_i \theta }})^A\Sigma_i^{AC} \Sigma_j^{CB}
(\D_j \theta)^B\nonumber\\
&=&\int d^3x~ ({\overline {\Sigma\cdot\D \theta }})^A (\Sigma\cdot
\D \theta)^A\nonumber\\
&=&\int d^3x~ {\bar \theta}^A \left[(\Sigma \cdot \D )^\dagger (\Sigma\cdot
\D)\right]^{AB}\theta^B 
\label{21}
\eeqar

A metric of the form
\beq
ds^2 = \int d^3x~ {\bar \theta}^A \theta^A
= \int d^3x~ \Tr (M^{\dagger -1} \delta M^\dagger ~\delta M M^{-1})
\label{22}
\eeq
is the Cartan-Killing metric for $U(2N, {\bf C})$ (for each spatial point)
and leads to the
Haar measure
$d\mu (M, M^\dagger )$ for $U(2N, {\bf C})$. By comparison with this
we see that the volume measure for the metric (\ref{21}) can be written as
\beqar
[dA] &=& \det \left[(\Sigma \cdot \D )^\dagger (\Sigma\cdot
\D)\right]~ d\mu (M, M^\dagger )\nonumber\\
&=& d\mu (M, M^\dagger )~\exp\left({\Gamma + {\bar \Gamma}}\right)\label{23}\\
\exp (\Gamma ) &=& [\det (\Sigma\cdot \D)]_{reg}\label{24}
\eeqar 
In equation (\ref{24}), we have explicitly indicated that the determinant
is to be evaluated with proper regularization.
The regularization should be such that $\Gamma +{\bar \Gamma}$ is gauge-invariant.
The volume element for the real part of $A^a_i$ is then given as
\beq
[d{\rm Re}A^a_i] = \int  e^{\Gamma + {\bar \Gamma}}~ d\mu (M, M^\dagger )
~\delta [\sigma\cdot\del M M^{-1} +{\rm h.c.}] ~\delta [ \Tr (t^a \sigma\cdot \del
M M^{-1}) - {\rm h.c.}] 
\label{25}
\eeq

The calculation of the volume thus involves several distinct steps. The first is
the calculation of the determinant $\exp( \Gamma +{\bar \Gamma})$; the second 
is the reduction of the Haar measure $d\mu (M, M^\dagger )$ by the elimination of
the set of gauge transformations and finally we have to address the question
of the constraints given by the $\delta$-functions.

The full determinant can be calculated by computing the determinants
of the Dirac-like operators $\Sigma\cdot \D$ and its adjoint and putting the results
together in a gauge-invariant way. 
The regulated form of the
determinant of $\Sigma \cdot \D$ can be written as
\beq
\Gamma_{reg}= \Tr \log \Sigma \cdot \D - {M_2\over M_2-M_1} \Tr \log (\Sigma\cdot
\D +M_1) + {M_1\over M_2-M_1} \Tr \log (\Sigma\cdot
\D +M_2)\label{26}
\eeq
where $M_1$ and $M_2$ are regulator masses. We will need to use two 
regulators of the Pauli-Villars type, with coefficients as given,
to eliminate all the
divergences. 

We can calculate the determinant by a series expansion in powers of
the gauge potential. The only unusual point is that the
simplification of the traces are more involved because the
$\Sigma$-matrices do not commute with the Lie algebra
of the
$\A$'s. Indeed if this were not so, the determinat would be trivial, 
apart from
possible anomalies, since $\A$ has the form $-\del M M^{-1}$.

The term quadratic in the potentials is given by
\beqar
\Gamma^{(2)}&=& {1\over 2}\int_{x,y} 
~ \Tr ( t^A \del_i M M^{-1})(x) ~\Tr (t^B \del_j M M^{-1})(y) ~
\int {d^3k \over (2\pi )^3 } e^{ik\cdot (x-y)}~\Pi^{AB}_{ij} (k)
\nonumber\\
\Pi^{AB}_{ij} (k) &=& -{i\over 16\pi} \left[ {M_2 \over M_2-M_1} ({\rm sgn} M_1) -
{M_1\over M_2-M_1} ({\rm sgn}M_2)\right] k_r \Tr ( [\Sigma_r , \Sigma_i T^A]
\Sigma_jT^B )\nonumber\\
&&\hskip .5in - {k\over 64} \left( \delta_{rs} + {k_r k_s \over k^2}\right) 
\Tr (\Sigma_r \Sigma_i T^A \Sigma_s \Sigma_j T^B)
\label{27}
\eeqar
where ${\rm sgn}M = {M \over \vert M\vert}$.
The first term in $\Pi^{AB}_{ij}$ corresponds to
a Chern-Simons term. The second term will be seen to be
similar to the one-loop vacuum polarization result in
three dimensions; the factor $\delta_{rs} + k_r k_s /k^2$ is correct with the
connecting plus sign, the usual projection operator will emerge once 
the traces are
evaluated.
The following two observations help to simplify these expressions.
First, notice that $\A_i $ obeys the identity
\beq
\del_i \A_j - \del_j \A_i +[\A_i ,\A_j ]=0
\label{28}
\eeq
so that, to the quadratic order in the potentials we have
$\del_i \A_j - \del_j \A_i \approx 0$. Secondly, the traces are in the adjoint
representation of
$U(2N)$, but these can be converted to the fundamental representation.
For example, using the completeness of the $t^A$'s, we can write
\beqar
[ T^A , \Sigma_r ]^{BC} &=& -i f^{ABD} \Sigma^{DC}_r + \Sigma^{BD} if^{ADC}\nonumber\\
&=& -\Tr ( [t^A, t^B] \sigma_r t^C ) - \Tr ( t^B \sigma_r [t^A,t^C])\nonumber\\
&=& \Tr ( t^B [t^A ,\sigma_r ] t^C )\label{29}
\eeqar
Using the algebra of the $\Sigma$'s, we then get
\beqar
\Tr [\Sigma_r , \Sigma_i T^A] \Sigma_j T^B &=&
2i \epsilon_{rik} \Tr ( \Sigma_k T^A \Sigma_j T^B) - \Tr \left( \sigma_i [t^A , \sigma_r]
\sigma_j \half i f^{C MN} t^C ) (-i f^{B MN} \right)
\nonumber\\
&=&2i \epsilon_{rik} \Tr ( \Sigma_k T^A \Sigma_j T^B) - {C_2\over 2}
\Tr \left( \sigma_i [t^A,\sigma_r]\sigma_j t^B\right)\label{30}
\eeqar   
where $C_2$ is the quadratic Casimir for the adjoint representation of
$U(2N)$, $f^{AMN}f^{BMN} = C_2 \delta^{AB}$. With $\epsilon_{rik} k_r \A_i \approx 0$,
the first term gives zero for the Chern-Simons contribution, reducing the trace to
the trace in the fundamental. Similar simplification can be done for all the other terms and
the final result is
\beqar
\Gamma^{(2)}&=& \half C_2 \Biggl[ -{i\over 16\pi}\left( {M_2 \over M_2-M_1} ({\rm sgn} M_1) -
{M_1\over M_2-M_1} ({\rm sgn}M_2)\right) \int \epsilon^{ijk} \del_i A^a_j ~A^a_k
\nonumber\\
&&\hskip .7in -{1\over 128} \int F^a_{ij} {1\over \sqrt{-\nabla^2}} F^a_{ij}
+{1\over 32} \int \phi^a \sqrt{-\nabla^2}~\phi^a \Biggr]
\label{31}
\eeqar
Here $F^a_{ij} \approx \del_i A^a_j - \del_j A^a_i$ to the order we have calculated.
The terms
involving
$\phi$'s are eventually set to zero by the constraints (\ref{9}, \ref{10}). 
The form of the $\phi$-terms could also change depending on the regulators, but
the final answer is unambiguous since we can set them to zero anyway.
The
Chern-Simons term will cancel out when we take $\Gamma +{\bar \Gamma}$, 
as it should, since
there is no  parity violation in pure (3+1)-dimensional gauge theory.
Using these simplifications, we get for the volume measure
\beqar
[d{\rm Re }A]&=& \int d\mu (M, M^\dagger )~\exp\left({\Gamma + {\bar \Gamma}}\right)
~\delta [\sigma\cdot\del M M^{-1} +{\rm h.c.}]\nonumber\\
&&\hskip 2in \times  ~\delta [ \Tr (t^a \sigma\cdot \del
M M^{-1}) - {\rm h.c.}]\nonumber\\
\Gamma + {\bar \Gamma} &=& -{C_2\over 128} 
\int F^a_{ij} {1\over \sqrt{-\nabla^2}} F^a_{ij}~+~ {\cal O}(A^3)
\label{32}
\eeqar
Based on gauge invariance, we can say that part of the higher order terms will
render the first term fully invariant, so that the result is of the form
\beq
\Gamma + {\bar \Gamma} = -{C_2\over 128} 
\int F^a_{ij} \left[{1\over \sqrt{-(\del +A)^2}}\right]^{ab} F^b_{ij}~+~ 
{\cal O}(A^3)
\label{33}
\eeq
with $F^a_{ij} = \del_i A^a_j - \del_j A^a_i +f^{abc} A^b_i A^c_j$.

We now turn to the Haar measure $d\mu (M, M^\dagger )$. We are interested in factoring out
the gauge transformations which act as $M^g = gM$, $g\in U(N)$.
Out of $M$ we can construct the gauge-invariant quantities
$H =M^\dagger M$ and $W_i = M^\dagger \sigma_i M$.
We write a generic $M$ as
\beq
M =\left[ \matrix{a&b\cr c&d\cr}\right]
\label{34}
\eeq
where $a, b, c, d$ are $(N\times N)$-matrices. We can take $a$ and $d$ to be invertible in
general \cite{gilmore}. Elements of the combinations $H$ and $W_i$ give
$a^\dagger a$, $d^\dagger d$, $a^\dagger d$, $c^\dagger d$, $b^\dagger d$, etc.
They can thus be regarded as functions of $H, W_i$. The square roots of
$a^\dagger a$ and $d^\dagger d$ can be defined by diagonalizing them.
We then see that we can write
\beqar
a= U~\sqrt{a^\dagger a}~,\hskip .3in && \hskip .3in b = U ~ \beta\nonumber\\
c= U~ \gamma, \hskip .6in && \hskip .3in d = U~ V \sqrt{d^\dagger d}
\label{35}
\eeqar
where $U$ and $V$ are unitary matrices;
$V$ is determined from $a^\dagger d$ as a
function of
$H, W_i$. Likewise $\beta$ and $\gamma$ are given by $c^\dagger d$ and
$b^\dagger d$. Thus the matrix $M$ can generally be parametrized as
\beq
M = \left[ \matrix {U&0\cr 0&U\cr}\right] ~N
\label{36}
\eeq
where $N$ is a function of $H, W_i$. The Haar measure is given by the top rank
differential form $dM M^{-1}\wedge  dM M^{-1}\cdots M^{\dagger -1} dM^\dagger
\wedge M^{\dagger -1} dM^\dagger \cdots$ where we substitute (\ref{36}) for
$M$. This brings out a factor $d\mu (U)= dU U^{-1}\wedge dU U^{-1}\cdots $
which is the volume of the gauge part, $U(N)$. The remainder is
given entirely in terms of the gauge-invariant combinations $H, W_i$.
In other words, we have $d\mu (M, M^\dagger )= d\mu (U) d\mu (H, W )$;
$d\mu (H, W )$ is the volume on the coset space $U(2N, {\bf C})/ U(N)$.
By taking the product of this formally over all spatial points, we have
\beq
d\mu (M, M^\dagger ) =  \prod_x ~d\mu (U) ~  \prod_x ~d\mu (H, W )
\label{37}
\eeq
The elimination of the gauge part of the measure is now trivial,
we just get $\prod_x ~d\mu (H, W )$.

Finally, it is easy to see that
the $U$-dependence of the constraints drops out from the $\delta$-functions
in (\ref{25}) or (\ref{32}); they can be
written in terms of $N$ or $H,W_i$. Combining results (\ref{32}) and (\ref{37})
and the arguments given above, we can write the gauge-invariant measure
as
\beqar
d\mu (\C )&=& \prod_x ~d\mu (H, W )~~\delta [\sigma\cdot\del N N^{-1} +{\rm
h.c.}] ~\delta [ \Tr (t^a \sigma\cdot \del
N N^{-1}) - {\rm h.c.}]\nonumber\\
&&\hskip 1in \times ~\exp\left( \Gamma +{\bar \Gamma}\right)\nonumber\\
&\equiv& d\mu~\exp\left( \Gamma +{\bar \Gamma}\right)\label{38} 
\eeqar
where $\Gamma +{\bar \Gamma}$ has the expansion (\ref{33}).

We can now write the  inner product for states $\vert 1\ra$ 
and $\vert 2\ra$, with the gauge-invariant
wavefunctions $\Psi_1$ and $\Psi_2$, as
\beq
\la 1\vert 2\ra = \int d\mu \exp\left( \Gamma +{\bar \Gamma}\right)~ \Psi^*_1 
\Psi_2
\label{39}
\eeq
The key result of this paper is this formula
for the inner product, along with (\ref{33}, \ref{38}),
which summarize our results on the gauge-invariant 
volume element for the
configuration space $\C$.
Notice that, as in the (2+1)-dimensional case, 
the term $\Gamma +{\bar \Gamma}$ is proportional 
to the quadratic Casimir 
$C_2$, which vanishes for the Abelian theory, once again
indicating a significant difference between the Abelian 
and nonabelian cases.

\section{Discussion}

Equation (\ref{39}) for the inner product
shows that the matrix elements of the 
(3+1)-dimensional theory
can be reduced to the correlators of a three-dimensional
Euclidean gauge theory with the action
$\Gamma +{\bar \Gamma}$ and functional measure $d\mu$.
We have obtained the quadratic terms in this action, but
not yet calculated the terms
which will involve gauge-invariant
combinations which are cubic and higher order in the fields,
although some of these higher terms can be inferred from
gauge invariance.
Nevertheless, it is still interesting to look ahead and see what 
implications our results may have for the physics of the 
gauge theory. 

We can establish some properties of the $(3+1)$-dimensional
theory by comparison with the
$(2+1)$-dimensional theory.
The vacuum wave function for the $(2+1)$-dimensional theory
was of the form $\exp \left[ - (\pi/2e^4N ) \int B^2\right]$ for long 
wavelength modes. With such a wave function, 
for the Wilson loop $W_F (C)$ in the  fundamental representation of 
$SU(N)$ we find
\begin{eqnarray}
\la W_F (C) \ra &=& {\rm constant}~~\exp \left[ - \sigma {\cal
A}_C \right]\nonumber\\
{\sqrt{\sigma }}&=& e^2 \sqrt{{N^2-1\over 8\pi}}
\label{40}
\end{eqnarray}
This result was obtained in the Hamiltonian description; nevertheless, based
on the full Euclidean invariance of the Wick rotated theory, this may be
expressed as
\beq
\int d\mu ({\cal C}) ~\exp \left( - \int {F^2 \over 4e^2}\right)
~W_F (C) = \la W_F (C) \ra = {\rm constant}~~\exp \left[ - \sigma {\cal
A}_C \right]
\label{41}
\eeq
This version may in turn be interpreted as the equal time correlator in the
$(3+1)$-dimensional theory with a vacuum wave function of the form
$\sim \exp (- \int F^2 /8e^2)$. Thus if the $(3+1)$-dimensional 
theory has a vacuum wave function
\beq
\Psi_0 \sim \exp \left( - \int {F^2 \over 8 \Lambda}\right)
\label{42}
\eeq
then we get confinement and a string tension
\beq
\sqrt{\sigma}= \Lambda \sqrt{{N^2-1\over 8\pi}}
\label{43}
\eeq
We therefore assume that the wave function has the form (\ref{42})
and ask what other implications it may have. The mass of a
$0^{++}$ glueball in the lower dimensional theory is given by
\beq
\la B^2(x) B^2 (0) \ra = \int d\mu ({\cal C})
~\exp \left( - \int {F^2 \over 4e^2}\right)
\sim \exp (- M_{0^{++}} \vert x\vert )
\label{44}
\eeq
for large separations $\vert x\vert$.
The mass $M_{0^{++}}= \alpha e^2N$, where $\alpha$ is, in
principle, calculable in  the Hamiltonian formulation of the 
$(2+1)$-dimensional theory.
An explicit calculation is difficult; lattice data show that
$\alpha \approx 0.808$ \cite{teper} as $N\rightarrow \infty$. We can also think of the result
(\ref{44}) as an equal time correlator in the $(3+1)$-dimensional theory,
for the wave function (\ref{42}) (with $e^2 \rightarrow \Lambda$),
in which case the glueball mass is given by 
$M_{0^{++}} = \alpha \Lambda N $. This means that, if the wave function
(\ref{42}) is a good description in the $(3+1)$-dimensional case,
the ratio $M_{0^{++}}/\sqrt{\sigma}$ is the same in the $(3+1)$-
and $(2+1)$-dimensional theories.
Collecting results
\beqar
{M_{0^{++}} \over \sqrt{\sigma}} &=& \alpha \sqrt {8\pi} {N \over \sqrt{N^2 -1}}\nonumber\\
&\approx&  4.05~{N \over \sqrt{N^2 -1}}\label{45}
\eeqar
where we have used the lattice value for the $(2+1)$-dimensional
theory. This is then a prediction, based on the premise of (\ref{42}), for the
$(3+1)$-dimensional theory. The lattice estimate of this quantity
for the $(3+1)$-dimensional theory is approximately $3.37$ as $N \rightarrow \infty$
\cite{lucini}; the
discrepancy is about $20\%$.
Thus equation (\ref{42}) may be considered to be
a reasonable ansatz for a first approximation to the wave function.
The important question is whether
we can we derive it by solving the Schr\"odinger equation;
this is under study.

The approximate
dimension-independence of the glueball masses has been noted 
before in the context of lattice values. In the context of using wave functions, 
an argument which has some similarity
 to ours has been given in \cite{samuel}. In the context of a parton mass 
for gluons, a similar observation has been made by Philipsen \cite{philipsen}.

There is another lesson from the (2+1)-dimensional case that we can
use. In three Euclidean dimensions
an action of the form $\int F^2/4\Lambda$ can generate a mass gap.
This is not yet the gap for the (3+1)-dimensional theory, but a
cutoff on modes of low momenta when
integrations are actually carried out using
a wave function of the form (\ref{42}). In turn this can generate a 
mass gap for the
(3+1)-dimensional theory in much the same way as the cut-off
on low momentum modes due to the measure factors
in the (2+1)-dimensional analysis
can lead to a gap \cite{KN}.

\section*{Acknowledgements}

VPN thanks Dimitra Karabali and Peter Orland for a critical reading of the manuscript.
This work
was supported in part by  NSF grant number PHY-0070883
and by a PSC-CUNY grant.


\begin{thebibliography}{99}

\bibitem{KN} D. Karabali and V.P. Nair, Nucl. Phys. {\bf B464} (1996) 135; 
 Phys. Lett. {\bf B379} (1996) 141; Int. J. Mod. Phys. {\bf A12} (1997) 1161.

\bibitem{KKN1} D. Karabali, Chanju Kim and V.P. Nair, Nucl. Phys. {\bf B524}
(1998) 661; some of this work has been reviewed by H. Schulz, hep-ph/9908527.

\bibitem{KKN2} D. Karabali, Chanju Kim and V.P. Nair, Phys. Lett. {\bf B434}
(1998) 103.

\bibitem{everybody} J. Goldstone and R. Jackiw, Phys.Lett. {\bf 74B} (1978) 81;
R. Jackiw and S. Templeton, Phys.Rev. {\bf D23} (1981) 2291; 
M.B. Halpern, Phys.
Rev. {\bf D16} (1977) 1798; 
{\it ibid.} {\bf
D19} (1979) 517;
R.P. Feynman, Nucl.
Phys. {\bf B188} (1981) 479; 
M. Bauer and D.Z. Freedman, Nucl. Phys. {\bf B450} (1995) 209;
F.A. Lunev,
Phys. Lett. {\bf B295} (1992) 99;
M. Asorey, Phys. Lett {\bf B349} (1995) 125; 
I. Kogan and A. Kovner, Phys. Rev. {\bf D51} (1995) 1948;
S. Das and S. Wadia, Phys. Rev. {\bf D53} (1996) 5856; 
O. Ganor and J. Sonnenschein, Int. J.
Mod. Phys. {\bf A11} (1996) 122.

\bibitem{witten} E. Witten, Commun. Math. Phys. {\bf 92} (1984) 455;
S.P. Novikov, Usp. Mat. Nauk. {\bf 37} (1982) 3;
D. Karabali, Q-H. Park,
H.J. Schnitzer and Z. Yang, Phys. Lett. {\bf 216B} (1989) 307;
D. Karabali and H.J. Schnitzer, Nucl. Phys. {\bf B329} (1990) 649.

\bibitem{gaw} K. Gawedzki and A. Kupiainen, Phys. Lett. {\bf 215B} (1988) 119;
Nucl. Phys. {\bf B320} (1989) 649;
M. Bos and V.P. Nair, Int. J. Mod. Phys. {\bf A5} (1990) 959.

\bibitem{teper} M. Teper, Phys. Rev. {\bf D59} (1999) 014512;
B. Lucini and M. Teper, Phys. Rev. {\bf D66} (2002) 097502.

\bibitem{VPN1} A.D. Linde, Phys. Lett. {\bf B96} (1980) 289;
D. Gross, R. Pisarski and L. Yaffe, Rev. Mod. Phys. {\bf 53} (1981) 43;
for a recent discussion, see, for example, V.P. Nair, in 
{\it TFT-98: Thermal Field Theories and
their Applications}, U. Heinz (ed.), hep-ph/9811469.

\bibitem{magmass} V.P. Nair, Phys. Lett. {\bf B352} (1995) 117;
G. Alexanian and V.P. Nair, Phys.
Lett. {\bf B352} (1995) 435 ;
W. Buchmuller and O. Philipsen, Nucl. Phys. {\bf B443} (1995)
47; O. Philipsen, in {\it TFT-98: Thermal Field Theories and
their Applications}, U. Heinz (ed.), hep-ph/9811469; F. Eberlein,
Phys. Lett. {\bf B439} (1998) 130; Nucl. Phys. {\bf B550} (1999) 303;
R. Jackiw and S.Y. Pi,
Phys. Lett. {\bf B368} (1996) 131; {\it ibid.} {\bf B403} (1997) 297; 
J.M. Cornwall, Phys.
Rev. {\bf D10} (1974) 500 ;  {\it ibid.} {\bf D26} (1982) 1453; 
Phys. Rev. {\bf D57} (1998)
3694.

\bibitem{karsch} F. Karsch {\it et al}, Nucl. Phys. {\bf B474} (1996) 217; 
F. Karsch, M. Oevers and
P. Petreczky, Phys. Lett. {\bf B442} (1998) 291;
A. Cucchieri, F. Karsch and P. Petreczky, Phys. Lett. {\bf B497} (2001) 80;
O. Philipsen, Phys. Lett. B521 (2001) 273; 
Nucl. Phys. Proc. Suppl.
{\bf 106} (2002) 242.

\bibitem{KKN3} D. Karabali, Chanju Kim and V.P. Nair, Nucl. Phys. B566 (2000) 331.

\bibitem{singer} I.M. Singer, Physica Scripta {\bf 24} (1980) 817;
I.M. Singer, 
Commun. Math. Phys. {\bf 60} (1978) 7.

\bibitem{mitter} 
P.K. Mitter and C.M. Viallet, Commun. Math. Phys. {\bf 79} (1981) 457;
Phys. Lett. {\bf 85B} (1979) 246; M. Asorey and P.K. Mitter, Commun. Math. Phys.
{\bf 80} (1981) 43; O. Babelon and C.M. Viallet, Commun. Math. Phys. {\bf 81} (1981) 515;
Phys. Lett. {\bf 103B} (1981) 45.

\bibitem{orland} P. Orland, hep-th/9607134.

\bibitem{gribov} V. Gribov, Nucl. Phys. {\bf B139} (1978) 1;
T. Killingback and E.J. Rees, Class. Quant. Grav.
{\bf 4} (1987) 357;
W. Nahm, in {\it Proceedings of the IV Warsaw Symposium on Elementary
Particle Physics}, Z. Adjuk (ed.) (Warsaw, 1981).

\bibitem{baal} P. Koller and P. van Baal, Ann. Phys. {\bf 174} (1987) 299;
Nucl. Phys. {\bf B302} (1988) 1; P. van Baal, Phys. Lett. {\bf 224B} (1989) 397;
Nucl. Phys. {\bf B351} (1991) 183; P. van Baal and N.D. Hari-Das, Nucl. Phys.
{\bf B385} (1992) 185; P. van Baal and B. van den Heuvel, Nucl. Phys.
{\bf B417} (1994) 215;
D. Zwanziger, Nucl. Phys. {\bf B209} (1982) 336; G. dell'Antonio
and D. Zwanziger, Nucl. Phys. {\bf B326} (1989) 333; Commun. Math. Phys.
{\bf 138} (1991) 291; D. Zwanziger, Nucl. Phys. {\bf B412} (1994) 657.
For recent updated views on this, see
T. Heinzl, hep-th/9604018; P. van Baal, hep-th/9711070.

\bibitem{gilmore} R. Gilmore, {\it Lie groups, Lie algebras and some of their
applications}, John Wiley and Sons, Inc., New York (1974). 

\bibitem{lucini} B. Lucini and M. Teper, JHEP 0106 (2001) 050.

\bibitem{samuel} S. Samuel, Phys. Rev. {\bf D55} (1997) 4189.

\bibitem{philipsen} O. Philipsen, Nucl. Phys. Proc. Suppl. {\bf 106}
(2002) 242; Nucl. Phys. {\bf B 628} (2002) 167. 
 

\end{thebibliography}
\end{document}